\documentclass[sigconf,10pt]{acmart}
\usepackage{multirow}
\usepackage{xspace}
\usepackage{listings}
\usepackage{subcaption}
\usepackage{tikz}
\usepackage{paralist}
\usepackage{cleveref}
\usepackage{tcolorbox}
\usepackage{xcolor}
\usepackage{amsthm}
\usepackage{thmtools}

\definecolor{gray}{HTML}{ededed}
\colorlet{shadecolor}{gray}
\tcbset{
	colback=gray,
	boxrule=0.5pt,
	boxsep=1pt,
	left=3pt,
	right=3pt,
	top=3pt,
	bottom=3pt,
	sharp corners,
	colframe=black,
}

\newcommand{\researchquestion}[1]{
	\begin{tcolorbox}
		#1
	\end{tcolorbox}
}

\newcommand{\headline}[1]{\vspace{+0.2cm}\noindent\textbf{#1}}

\declaretheoremstyle[
    spaceabove=6pt, 
    spacebelow=6pt,
    headfont=\normalfont\bfseries,
    notefont=\mdseries, 
    notebraces={(}{)},
    bodyfont=\normalfont,
    postheadspace=1em,
    headpunct={:},
    headformat=\NAME\NOTE,
]{defstyle}

\title{I've Got 99 Problems But FLOPS Ain't One}

\author{Alexandru M. Gherghescu\textsuperscript{1} \hspace{1em} 
Vlad-Andrei Bădoiu\textsuperscript{1} \hspace{1em} 
Alexandru Agache\textsuperscript{1} \hspace{1em} 
Mihai-Valentin Dumitru\textsuperscript{1} \hspace{1em} 
Iuliu Vasilescu\textsuperscript{1} \hspace{1em} 
Radu Mantu\textsuperscript{1} \hspace{1em} 
Costin Raiciu\textsuperscript{1,2}}

\def \authors{Alexandru M. Gherghescu,
Vlad-Andrei Bădoiu,
Alexandru Agache,
Mihai-Valentin Dumitru,
Iuliu Vasilescu,
Radu Mantu,
Costin Raiciu}

\affiliation{
  \institution{
    \textsuperscript{1}University Politehnica of Bucharest
    \hspace{1em} \textsuperscript{2}Broadcom Inc.
  }
  \city{}
  \country{}
}

\begin{abstract}

\noindent Hyperscalers dominate the landscape of large network deployments,
yet they rarely share data or insights about the challenges they face. In
light of this supremacy, what problems can we find to solve in this space? We
take an unconventional approach to find relevant research directions,
starting from public plans to build a \$100 billion datacenter for
machine learning applications~\cite{100bds}. Leveraging the language models
scaling laws, we discover what workloads such a datacenter might carry
and explore the challenges one may encounter in doing so, with
a focus on networking research. We conclude that building the datacenter and
training such models is technically possible, but this requires novel wide-area transports for inter-DC communication,
a multipath transport and novel datacenter topologies for intra-datacenter communication, high
speed scale-up networks and transports, outlining a rich research agenda for the networking community.

\end{abstract}

\begin{document}

\acmYear{2024}\copyrightyear{2024}
\acmConference[HOTNETS '24]{The 23rd ACM Workshop on Hot Topics in Networks}{November 18--19, 2024}{Irvine, CA, USA}
\acmBooktitle{The 23rd ACM Workshop on Hot Topics in Networks (HOTNETS '24), November 18--19, 2024, Irvine, CA, USA}
\acmDOI{10.1145/3696348.3696893}
\acmISBN{979-8-4007-1272-2/24/11}

\renewcommand{\shortauthors}{Gherghescu et al.}

\maketitle

\section{Introduction}

The past few years have witnessed a revolution in the field of natural language
processing, with large language models (LLMs) emerging as extraordinary tools
for a wide range of language tasks (summarization, translation, classification,
natural language inference, question answering, text
generation~\cite{vaswani2017attention,devlin2018bert,lewis2019bart,brown2020language,raffel2020exploring,liu2019roberta,beltagy2020longformer}).
The underlying assumption is that the number of parameters of a model is the
single biggest indicator of model quality on a diverse range of
benchmarks~\cite{zero,megatron,shoeybi2019megatron,smith2022using}. It is then
understandable that many companies invest resources in training such models. As
public models reach hundreds of billions of
parameters~\cite{smith2022using,brown2020language,gopher,dubey2024llama,chowdhery2023palm},
and closed models like ChatGPT presumably surpass the trillion-parameter
threshold~\cite{gpt4_arch}, scaling
laws~\cite{deepmind_scale,openai_scaling} predict no immediate end to this
growth. We are in the midst of a race to develop ever-larger models, with
datacenter computing at the heart of this competition.

Training language models only requires matrix multiplications, but the
sheer size of these matrices makes compute a valuable
resource. Accelerators (GPUs, TPUs) have emerged as the
de-facto standard of model training,
dominating both compute and network traffic generation in AI-focused
datacenters.  Clusters comprising tens of thousands of accelerators
are used currently for training
Transformer-based~\cite{vaswani2017attention} models. Distributed
training frameworks~\cite{fsdp,megatron,zero} have been developed to
mask network bottlenecks and keep the accelerators busy during model training.  It is well
understood today that the network is the primary bottleneck, as various
works~\cite{megatron,touvron2023llama,reducing_activation,jiang2024megascale,rajbhandari2021zero,chowdhery2023palm}
have reported being able to use accelerators at only 40-60\% capacity
when scaling-out, due to the inherent limitations present in
large-scale model training (synchronization points, large number of
peers etc.).

\textit{\textbf{What are the main challenges in building a \$100B datacenter?}}
Foremost among these is cooling and securing sufficient electrical power, as we
anticipate consumption on the order of gigawatts — a scale not readily
available in a single location within the United States. 
Once these fundamental infrastructure challenges are addressed,
we encounter the primary technical impediment: training scalability. As we
approach systems with millions of GPUs, several critical questions arise: How
will training methods adapt to such massive scale? What will be the upper
limits of model size? How will networking requirements evolve to support this
unprecedented level of computation and data movement?

This study provides insights into the future of large-scale AI infrastructure
and the technical hurdles that must be overcome to realize the next generation
of massive language models, with the primary focus on model training.

\section{Infrastructure}
\label{section:infrastructure}

We base our calculations on the recently announced Blackwell GB200 GPU from
Nvidia~\cite{nvidiab200_1,nvidiab200_2,nvidiab200_3}, with novel NVL72 (72~GPUs
/ rack) packaging~\cite{nvidia2024nvl72, 10665247}. 
The NVL72 racks cost \$3M, have a TDP of 130~kW in full load~\cite{dcs2024,
gtc2024}, and can output 1,440~PFlops in sparse FP4 computation (half of that
for dense operations). Each GPU has 192~GB of memory and NVLink speeds of
14.4~Tbps (the so called \emph{scale-up network}), with a 800~Gbps NIC connected to the \emph{scale-out network}.

Following current trends in data center budgeting~\cite{obin2024whomakes}, an
allocation of 70\% of the budget to compute seems reasonable, giving us an
approximative 23.3K racks containing \textbf{1.67M GPUs} with a maximum power
consumption of about \textbf{3~GW} and \textbf{16,800,000~PFlops} of FP4
performance for dense operations (no sparsity)
. That is, the datacenter can sustain $16.8e21$ FP4 floating point
operations per second in full load.



Assuming that storage and networking require an additional fifth of the power
needed for compute~\cite{obin2024whomakes}, we calculate a total power
consumption of 3.6~GW for IT equipment.
Combined with Microsoft's reported Power Usage Effectiveness (PUE) numbers,
mainly between 1.15 and 1.3~\cite{msdspue2024}, the total power consumption of
the data center would end up between the ranges of 4.16~GW and 4.71~GW; this is consistent
with public reports about OpenAI datacenter plans \cite{oai5gw}. Where can
we find this much energy in the US?

One option is to build new generation capacity, but this takes time. The easier option
is to use existing spare capacity in the near future. 
We surveyed US grid operators~\cite{eia} and identified the maximum disposable power for
all grids by computing the difference between the max registered power
generation and max registered power demand in the last two years, with the most
promising shown in Table~\ref{tab:balancing_authority}.




\begin{table}[h]
\centering
\footnotesize
\begin{tabular}{|l|r|r|}
\hline
  \textbf{Balancing Authority} & \textbf{Max Available (MW)} & \textbf{Region} \\
\hline
  PJM  & 9915 & Mid-Atlantic \\
  SRP  & 2634 & Southwest \\
  NEVP & 2209 & Northwest \\
  BPAT & 2143 & Northwest \\
\hline
\end{tabular}
  \caption{Maximum available energy for the top energy producer grids in the US.}
\label{tab:balancing_authority}
\end{table}

While PJM is a viable candidate, it covers multiple states. To identify a
single geographic point, we grouped power sources within a specific radius into
power nodes. We then connected these nodes with vertices representing
high-voltage transmission lines. We excluded nodes where coal dominates energy
production, as coal plants are being phased out
gradually~\cite{coal_phase_out}. Additionally, considering that nuclear power
plants require downtime for maintenance and refueling~\cite{eia_nuclear},
there needs to be enough interconnect capacity the size of the largest producer in
each node to ensure uninterrupted datacenter operation. Following this
approach
, we couldn't find a single location meeting all requirements. However,
expanding our search to a 100 km radius revealed several suitable options near
Washington city. Consequently, from a power perspective we are constrained to
divide the datacenter into multiple units. This is consistent with news from
Microsoft \cite{msft-multipledc} and chinese hyperscalers \cite{china-multipledc} that distributed
training across datacenters is needed, and indeed possible.




Given the need to divide the datacenter, we propose an east-west coast split
due to several advantages. This geographical separation allows us to tap into
diverse renewable energy sources rather than relying on the sources available
in a single region. It also enhances fault tolerance against both natural and
man-made disasters, aligning with the east-west division of the US power grid.
Additionally, having datacenters on both coasts could provide low-latency
inference capabilities to users in both regions after the initial training
phase. For the west coast unit, a suitable location could be in the Northwest
regions, close to the Pacific DC Intertie powerline.
We have not explored
splitting the datacenter into three or more locations, leaving such analysis
for future research.

\section{Model scaling}


What model can we train using the above infrastructure? We start with the
classical Transformer model~\cite{vaswani2017attention}, since it is
the most studied and understood architecture, with the fairly standard
architectural variations
(RoPE~\cite{su2024roformer},
GeLU activation~\cite{hendrycks2016gaussian}, pre-layer
normalization~\cite{brown2020language}, no shared
embeddings~\cite{brown2020language,touvron2023llama,touvron2023llama2}, full
attention heads~\cite{brown2020language}, no
biases~\cite{brown2020language,touvron2023llama,touvron2023llama2}), in order to
draw 
bounds on model size. We later explore Mixture of Experts
as well. We ignore linear attention and sub-quadratic models~\cite{retnet,mamba,rwkv} in this
work, due to existing concerns for benchmark and long-range dependencies performance, and also 
because the training time should be similar.

\subsection{Scaling laws}

Given the estimated compute budget found in
section~\ref{section:infrastructure}, we proceed to determine suitable model
sizes. We use true FP4 arithmetic precision supported by hardware, instead of
quantization (on-the-fly decoding at compute time), in an attempt to maximize
device compute and memory utilization. We leave such approaches to future
research, as we're mainly concerned with networking, but note that there have
been avenues studying lower-precision
training~\cite{fishman2024scalingfp8trainingtrilliontoken,sun2020ultra,wang2023bitnet,ma2024era}.


\begin{figure*}
  \centering
  \begin{minipage}{0.38\textwidth}
    \includegraphics[width=\linewidth]{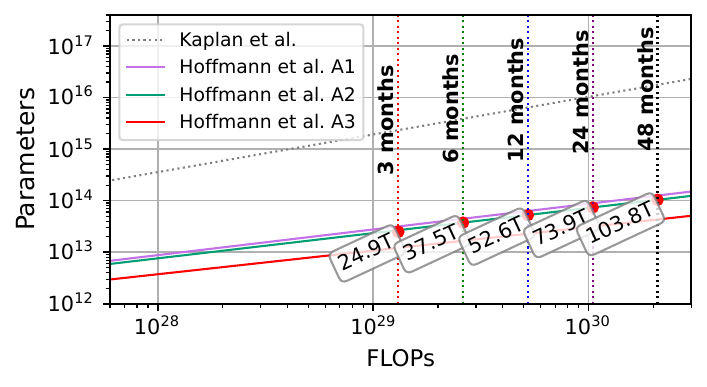}
    \caption{Loss-optimal model sizes}
    \label{fig:scale_law_extended}
  \end{minipage}
  \hfill
  \begin{minipage}{0.61\textwidth}
    \includegraphics[width=\linewidth]{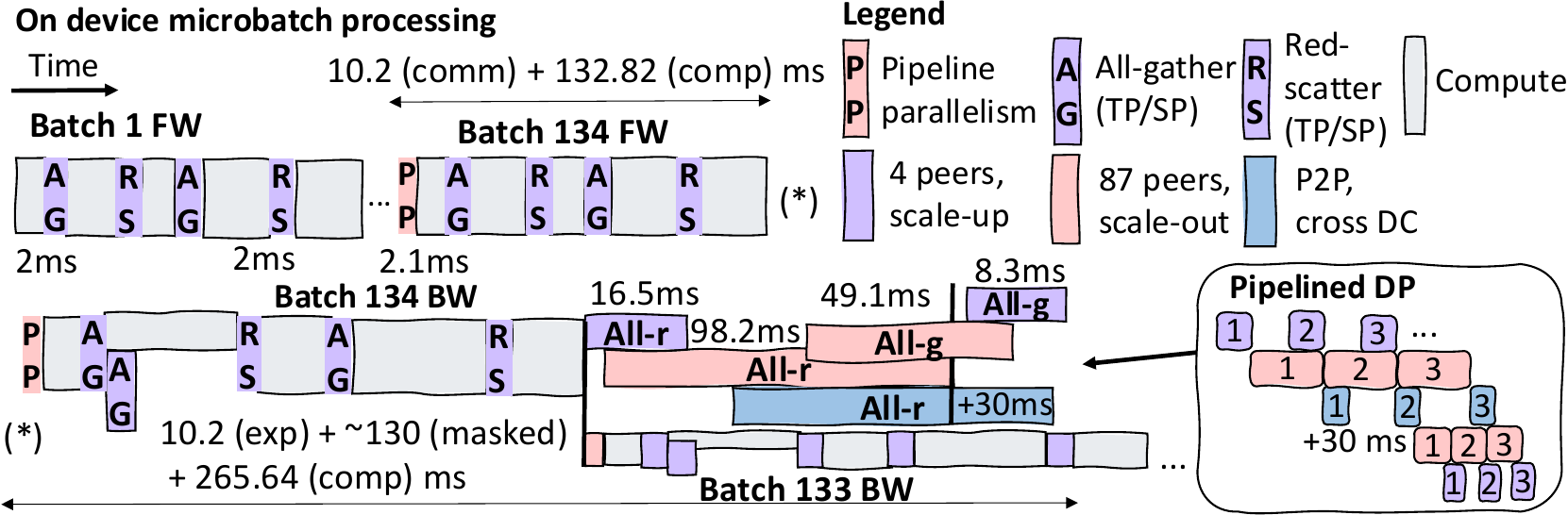}
    \vspace{-0.25cm}\caption{3D parallelism: per device forward+backward zoom-in}
    \label{fig:diag_zoom}
  \end{minipage}
\end{figure*}



\paragraph{\textbf{Training time.}} Based on scaling laws shown in Fig.~\ref{fig:scale_law_extended}, assuming no restrictions
on interconnect or bandwidth and full GPU utilization, we select a few
suitable model sizes that can be trained on our target datacenter. We
look at both Kaplan et al.~\cite{openai_scaling} and the 3 approaches from
Hoffmann et al.~\cite{deepmind_scale}, and choose Hoffmann et al.'s
approach number 2 (see Fig.~\ref{fig:scale_law_extended}).
%
%
%
We have examined in detail both 50T and 100T models, reaching qualitatively similar conclusions.
For brevity, we use the 103.8T parameter model in the rest of our paper.
For now we ignore imperfect networking, non-ideal operating
conditions, memory utilization, GPU arithmetic intensity and other such factors
that reduce GPU utilization and inevitably increase training time. 
We cover the effects of imperfect networks in \S\ref{sec:networking}.

\subsection{The architecture of the model}

For the 100T model size, we follow previous work on ideal depth-to-width
ratio~\cite{levine2020depth}. We also make several assumptions, as we are in
uncharted territory:

    \paragraph{\textbf{Vocabulary size of 256,000 tokens.}}
    Crucially, this helps capture multi-lingual data, as well as most of the
    Unicode characters. Models such as \textit{Gemma}~\cite{team2024gemma}
    or \textit{PaLM}~\cite{chowdhery2023palm} have popularized this size.

    \paragraph{\textbf{134 layers, 256 attention heads
    and 244,224 hidden size},} 
    totaling 96 trillion parameters ("100T" from now on).


    \paragraph{\textbf{Sequence length of 32k tokens.}} 
    The memory size of the self-attention layer scales quadratically
    with context length, limiting the amount of tokens that can be processed
    at once.
    A number of
    techniques~\cite{longrope,longlora,yarn,soaring,dubey2024llama} can be
    used to extend the context length after pre-training,
    mitigating this problem.


%

\section{Model Materialization}


Until now, we've built a theoretical model, which we move to materialize in this
chapter. We start from \textit{3D parallelism}~\cite{megatron},
as it's been shown to scale better for larger
Transformer models compared to techniques such as \textit{ZeRO-DP}~\cite{zero}
or \textit{FSDP}~\cite{fsdp}. We'll distribute the layers of the model to GPUs
in such a way as to keep tensor parallelism computation local (inside the NVLink
scale-up domain), while trying to minimize or overlap the communication times of
pipeline parallelism and data parallelism, which happen over the scale-out
800~Gbps links.

The total aggregate memory needed to train the model, 384.14~TB, is composed of
the memory for the model itself (144.04~TB for the FP8 master model,
computation of forward / backward passes in FP4 precision), gradients (48.01~TB),
optimizer states (192.05~TB), as well as residual memory from activations, which
we detail next.




Because 1 layer cannot fit inside a single GPU, we partition the weights and
activations with the scheme introduced
by~\cite{reducing_activation} (i.e. data, pipeline and tensor/sequence
parallelism, as well as selective activation recomputation). We partition the
activations across multiple devices, with no overlaps. We calculate activations
in bytes, per device, as $sbh(\frac{10}{t})$, where $s$ is the sequence length,
$b$ is the batch size and $h$ is the hidden dimension. In order to save device
memory, we offload the activations for $p$ pipeline parallel
stages~\cite{megatron} to host
memory~\cite{rajbhandari2021zero,zero,ren2021zero}, and bring them back in the
backward pass.
We determine the tensor parallel size $t$ by manually trying all 72 values (as
there's only 72 devices per rack), and picking the smallest $t$ that can hold
the model and active activations memory, arriving at $t=18$ (4 layers per
rack, \textasciitilde160~GB per device and 696 total data parallel replicas).




Importantly, we note that the layers inside a rack are all \textit{the same
layer} of different model replicas, as opposed to consecutive layers of the same
model replica. We show in section~\ref{section:communication} why we chose this
approach.



\paragraph{\textbf{Computation.}} With the above partitioning, we calculate the
theoretical ideal computation time per layer using the formulas
from~\cite{deepmind_scale,megatron,reducing_activation} (we assume the backward
pass is twice the forward pass), and obtain $132.82~ms$ for the forward pass (a
mere $0.97\%$ increase in computation compared to no activation recomputation).




\subsection{Network Communication}
\label{section:communication}

With 4 layers per rack, we have
18-way tensor/sequence parallelism, 134-way pipeline parallelism, and
696-way data parallelism (348-way per DC). Fig.~\ref{fig:diag_zoom} captures
the computation and overlapped communication during the forward and backward
passes for 1 device holding a chunk of a layer.

\paragraph{\textbf{Tensor and Sequence Parallel (TP/SP).}} As per~\cite{megatron}, inside
a layer, tensor/sequence parallelism needs to run two all-gather and two
reduce-scatter operations per forward pass, as there are 4 synchronization
points (the outputs of the attention layer, the feed-forward
layer, and the 2 layer norms), for a tensor of size $sbh$.
The resulting communication overhead is given by $2 \times sbh \times r
= 7.82~GB$, where $r$ is precision (0.5 bytes for FP4), and results in
$8.2~ms$ over the 1.8~TBps NVLink network (we approximate 2 all-gathers and 2
reduce-scatters as 2 all-reduce with a ring implementation). This communication
cannot be overlapped with computation, as tensor/sequence parallelism
introduces output-input dependencies between sub-layers (the attention, feed
forward and the 2 layer norms, as mentioned above). A similar communication cost
is incurred for the backward pass.




\paragraph{\textbf{Pipeline Parallel (PP).}} Between racks, the communication for
pipeline parallelism reduces to a point-to-point (P2P) exchange, for both forward and backward. Because of sequence
parallelism~\cite{megatron}, each device only needs to send its chunk of the
input to the device with the same rank in the next layer (such a chunk has
$\frac{sbh}{t} \times r = 217~MB$). This leads to a communication time of $2.17~ms$ over
the 800~Gbps scale-out network (once for forward, once for backward). Similarly
to tensor parallelism, this communication is exposed,
since it introduces dependencies between 2 consecutive layers.



\paragraph{\textbf{Data Parallel (DP).}} After exchanging the data as described
above, we need to run all-reduce operations in the backward pass to synchronize
gradients between model replicas.

To keep communication time low, especially for the gradient all-reduce
in the backward pass, we use a hierarchical data-parallel approach. Since
we partitioned the model to have 4 similar layers of different model replicas in
a single rack, we outline a 5-stage approach to pipeline this DP communication.

The first step is to all-reduce the gradients for the 4 layers inside a rack.
With gradients of 358~GB per layer, this takes $16.57~ms$. The second step
is an all-reduce inside a datacenter. In order to avoid extra communication,
each layer in a rack (out of the 4) only exchanges $1/4th$ of its gradients with
other racks. At 89.5~GB per chunk, this takes $98.27~ms$ over the scale-out
network, with 87 peers per DC. In the third step, each chunk needs to be
all-reduced with its corresponding chunk from the other DC (this is
just communication between pairs of 2 devices, with gradients of size
$\frac{89.5~GB}{18~devices \times 87~racks} = 57~MB$ per device, with roughly
800K devices per DC), with an RTT of $60~ms$. An all-gather in the fourth step
in the scale-out network, between the 87 peers in a DC, reconstructs
the gradient chunks (the $1/4th$-sized chunks per layer), with a time of
$49.14~ms$, followed by a scale-up all-gather, to finally "merge" back
the chunks into the layers, with a time of $8.3~ms$. Although this seems like a
lot of time, these communication stages can be easily pipelined, as
they happen over different networks.
Instead of fully committing the device gradients for each of the 5 communication
steps outlined above, only small batches of data can be pipelined at a time
through the steps, facilitating network overlap for the 3 different networks
(scale-up, scale-out and cross-DC). See figure~\ref{fig:diag_zoom} for a visual
explanation.

In practice, each of the 72 devices in a rack operates mostly independently of
the others. In the first step, 1 device interacts with 3 others in the same rack
(as there are 4 layers). In the second step, 1 device interacts with 86 devices
in 86 other racks, per DC (devices with the same rack rank), which hold the same
layer replica. In the third step, there is pair-wise communication between DCs.
In the fourth step, another communication between 87 peers happens, and finally
in the last step 1 device interacts with 3 devices in the scale-up network,
similarly to the first step.

\section{Mixture of Experts}

We also study scaling to 100T parameters using the Mixture of Experts
architecture~\cite{mixture-of-experts,mixtral,fedus2022switch}.
In line with previous work~\cite{mixtral}, we choose 8 experts per feed-forward
layer, and $top_k=2$ active experts to which tokens are routed per
pass.




\paragraph{\textbf{Model architecture.}} We choose to keep the total number of
parameters similar (100T), but have 8 experts per feed-forward layer. This
results in an 8x17T model, with \textit{$118$ layers}, \textit{$256$
attention heads} and \textit{$109,568$ hidden dimension}. We use the same
precision, vocabulary size and context length, 
as well as router networks before each feed-forward layer. As the number of
total parameters is similar, the memory required for the model amounts to
roughly the same value (385.54~TB).
%
We similarly keep 4 layers per rack (18 devices per layer). Due to less layers
for the MoE model, we have 788 total replicas, and \textasciitilde181~GB per
device.




\paragraph{\textbf{Computation and communication}} change from the dense
Transformer. Since the forward and backward passes are routed through only 2
experts from the 8 total,
the MoE model acts as a $28.4T model$
(only \textit{$28.4T$ parameters are active} during computation). This yields a
computation time of $45.18~ms$ per forward pass and twice per backward, per
layer. Communication time also changes due to the lower number of layers and
hidden dimension. Tensor/sequence parallelism, as well as pipeline parallelism,
have to communicate inputs of size $1.75~GB$, resulting in
$3.68~ms$, and $0.97~ms$, respectively, per forward pass, per layer. Similarly
to the dense Transformer, this cannot be overlapped. The communication pattern
for data parallelism doesn't change. Although only 2 experts are routed per
forward and backward pass, all experts need to have their
gradients synchronized with other replicas. With a total of $408~GB$ of
gradients per layer to exchange, the 5 stages detailed in
chapter~\ref{section:communication} last, respectively: $18.9~ms$ scale-up
all-reduce, $112.23~ms$ scale-out all-reduce with 98 total peers, cross-DC
pair-wise communication, $66.1~ms$ scale-out all-gather and $9.4~ms$ scale-up
all-gather.

\paragraph{\textbf{Routing network.}} In order to avoid extra communication of
the router, we split the 8 experts on all of the 18 devices (each device gets a
chunk of $1/18th$ of each of the 8 experts), in regular tensor parallel fashion.
Since the router is not big, we additionally save communication in the forward
pass by replicating the router on each device. There is extra communication in
the backward pass for the router's gradient, but the size is very small
compared to the other model parameters. This sharding strategy should also help
balance computation on all devices, irrespective of expert chosen.



\paragraph{\textbf{Implications.}} The backward computation time is less
than the DP gradient exchange,
even with an ideal network, suggesting we are now in a network-bound regime, as
opposed to compute-bound, as was the case for the dense Transformer.

\begin{figure*}
  \centering
  \begin{subfigure}{0.33\textwidth}
    \includegraphics[width=\linewidth]{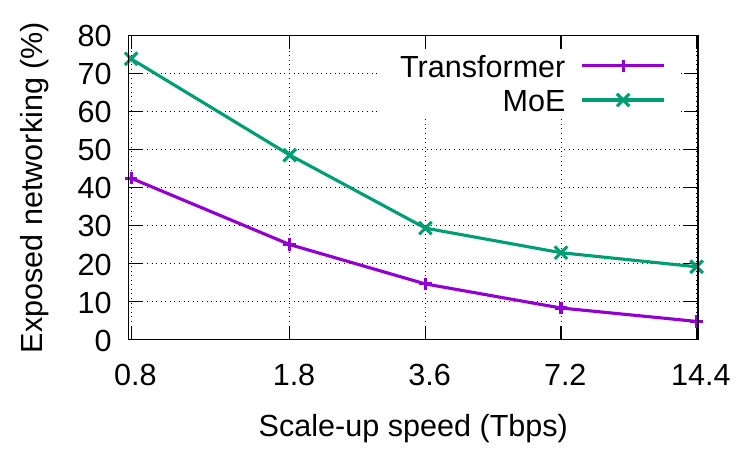}
    \caption{Effect of scale-up speeds}
    \label{fig:scale-up}
  \end{subfigure}
  \hfill
  \begin{subfigure}{0.33\textwidth}
    \includegraphics[width=\linewidth]{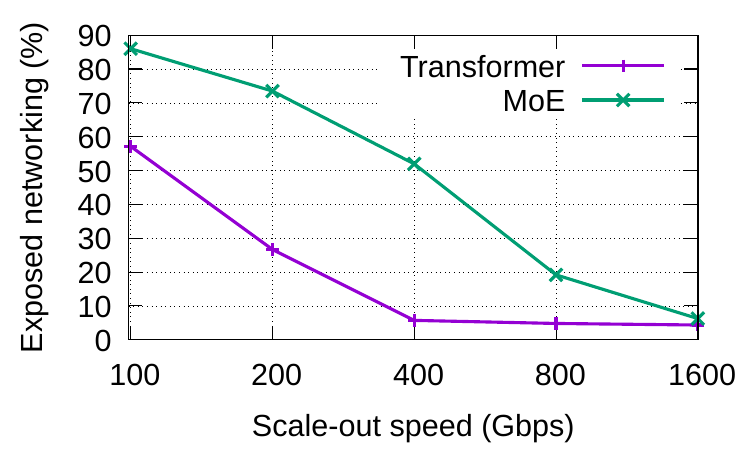}
    \caption{Effect of scale-out speeds}
    \label{fig:scale-out}
  \end{subfigure}
  \hfill
  \begin{subfigure}{0.33\textwidth}
    \includegraphics[width=\linewidth]{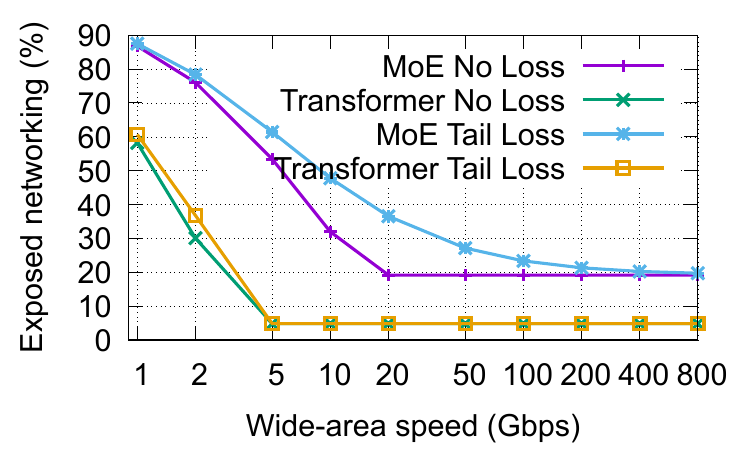}
    \caption{Effect of wide-area speeds}
    \label{fig:distributed}
  \end{subfigure}
  \caption{Effect of network capacity on exposed communication.}
\end{figure*}

\section{Network topology and transport protocols}
\label{sec:networking}

The time required to finish a step in the distributed model training includes
the compute and the time waiting for the network, the so-called exposed
communication time.

Our predictions show that, for each batch of training data, the GPUs will take
approximately $132~ms$ to perform the calculations for the forward pass, and
$264~ms$ for the backward pass for the standard model, and $45$ / $90~ms$ for
the Mixture of Experts model. Much of the networking is overlapped with
computation, and thus hidden.
The scale-out communication for pipeline parallelism, and the scale-up
communication for tensor/sequence parallelism is however exposed; the
scale-out all-reduce for the MoE model is also partly exposed. In our
analysis we ignore pipeline effects---namely exposed communication at the
start and end of each training batch---and only focus on the steady state,
where all GPUs are busy (see Fig. 4 in \cite{megatron}).

Our model predicts that exposed networking takes 5\% for the standard
transformer and 20\% for MoE, per training step, with 14.4~Tbps scale-up and
800~Gbps scale-out capacity, non-blocking networks and optimal transport
protocols. For MoE, the data-parallel all-reduce takes $112~ms$ of which
$90~ms$ are masked. In general, training MoE models is more network
intensive, and this is reflected in more exposed networking time. 

How do scale-up and scale-out speeds affect exposed networking?
Figure~\ref{fig:scale-up} shows how different scale-up speeds affect exposed
communication time when scale-out speed is 800Gbps. When scale-up speeds are
the same as scale-out (800Gbps), exposed time is more than 40\% for the
standard transformer and 75\% for the MoE version. Increasing the speeds to
multi-terabit dramatically reduces exposed networking to 5-20\%, highlighting
the importance of the scale-up network for distributed model training.

Figure~\ref{fig:scale-out} shows the effect of different scale-out networking
speeds while keeping the scale-up speed at 14.4~Tbps per GPU. Using just
100Gbps per GPU increases exposed networking to 55-85\%, mostly because of the
data-parallel communication across racks.  Increasing the scale-out speed beyond
400Gbps, does not reduce exposed networking time for the standard transformer
but it does help MoE; using 1.6~Tbps scale-out brings MoE exposed networking to
less than 5\%. 

Finally, figure~\ref{fig:distributed} shows how the provisioned wide-area
capacity, per GPU, affects exposed networking. If we have a lossless ideal
transport and at least 20~Gbps of wide-area capacity per GPU, the $30~ms$
propagation delay between the east and west coast datacenters is completely
masked by compute. However, if we have tail loss that inflates the tail FCT by
one RTT (60ms), the networking becomes exposed, especially when provisioned
capacity per GPU is lower. When retransmissions are required, especially for MoE
(backward compute just $90~ms$), wide-area networking becomes exposed unless we
have very high wide-area speeds.





\subsection{Building the scale-out network}
How do we connect 800K GPUs with 800Gbps NICs in each of our two datacenters?
The largest switches on the market today have 51.2~Tbps bisection bandwidth, and
can be configured with 64 800Gbps ports.  To build a fully provisioned Fat
Tree \cite{fattree}, we need 4 tiers with a total of 87.5K switches and 2.4 million
switch-to-switch links. Most of these links are tens of meters in length, which
makes cheap, passive DAC cables infeasible; optical transceivers cost 1K USD
even at 400~Gbps, so the estimated price tag for the wires alone is ~\$5 billion.
A common budget for networking is 10\% of the
datacenter cost~\cite{obin2024whomakes}, so \$5 billion per datacenter; our standard
Fat Tree design costs much more - we thus need cheaper alternatives, and the key
is to reduce the number of tiers.

\researchquestion{What is the most cost-effective network for
ML clusters that does not increase exposed networking time?}

\noindent Notwithstanding new research, there are two approaches which
can be used: increasing the effective switch radix with multiple
planes~\cite{multiplane} and reducing the number of endpoints per
network with independent rails~\cite{multirail}. We discuss these
below.

\headline{Using multiple-planes} is a way to achieve higher effective switch
radix with the same switching chips~\cite{multiplane}. The same 51.2T chip can
be also configured as a 128x400Gbps, 256x200Gbps or 512x100Gbps ports switch.
It's important to note that an 800Gbps link is in fact built using 8 parallel
100Gbps serdes blocks transmitting over the same wire. We can thus have the
same wire carry 4 or 8 lower-speed links for the same total capacity of
800Gbps.

To have four planes, we will build switches that contain four 51.2T ASICs and have 256
front panel ports. Inside the switch, each front panel port is divided into four
200Gbps links each of which is connected to a separate switching ASIC via
backplane copper traces. The total NIC bandwidth is still 800Gbps, but it is
obtained by connecting to 4 separate networks (planes) which use the same wires
and switches but are otherwise independent. Switch-to-switch links follow the
same pattern, where 4 200Gbps links share a wire.

With 256-port switches, three tiers are (more than) sufficient to connect all
the GPUs. Multi-plane thus reduces the number of switches and switch-to-switch links
by a third compared to the standard Fat Tree. Another significant benefit is
that it enables better traffic locality, as we now have four times more GPUs
connecting to the same TOR switch.

\headline{Using multiple rails.} How should we interconnect the GPUs to the
scale-out network? Current cloud network practice~\cite{jupiter} is to connect
all GPUs in the same rack to the ToR switch.  However, with an ML network, these
GPUs are already connected via the faster scale-up network, so the scale-out ToR
would carry little local traffic. Another observation is that GPUs from
different racks typically communicate in patterns, e.g. GPU 1 from rack 1 would
communicate with GPU 1 from rack 2 during data-parallel training.

\begin{figure*}
  \centering
  \begin{minipage}{0.33\textwidth}
    \includegraphics[width=\linewidth,trim=1cm 8cm 10cm 2cm,clip=true]{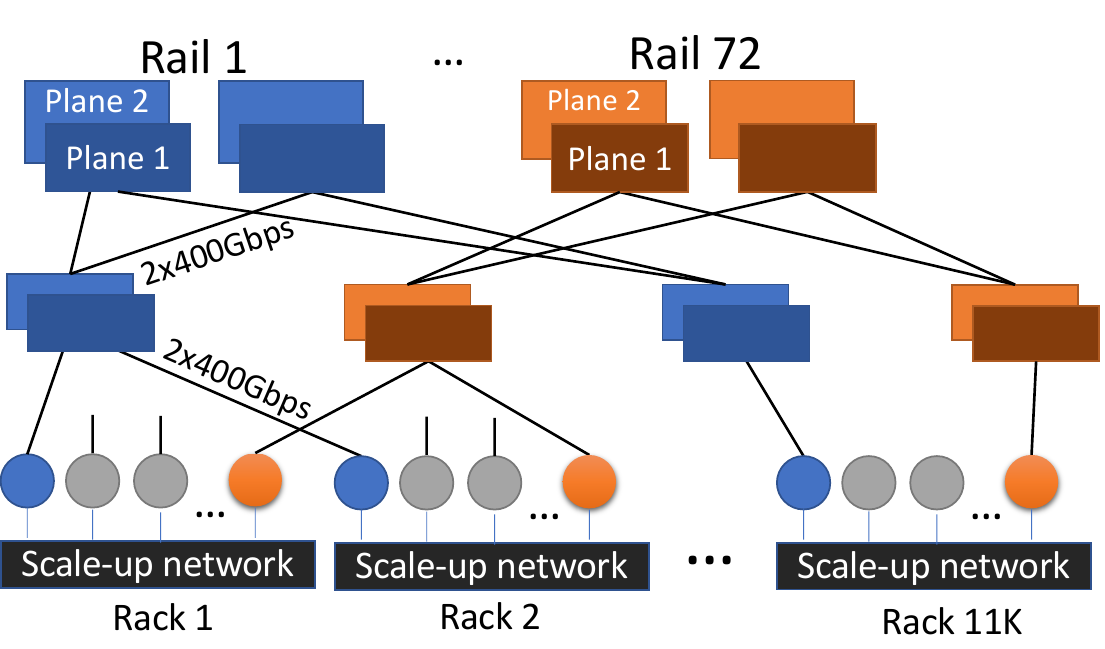}
    \caption{Possible topologies for AIML}
    \label{fig:topologies}
  \end{minipage}
  \hfill
  \begin{minipage}{0.33\textwidth}
    \includegraphics[width=\linewidth]{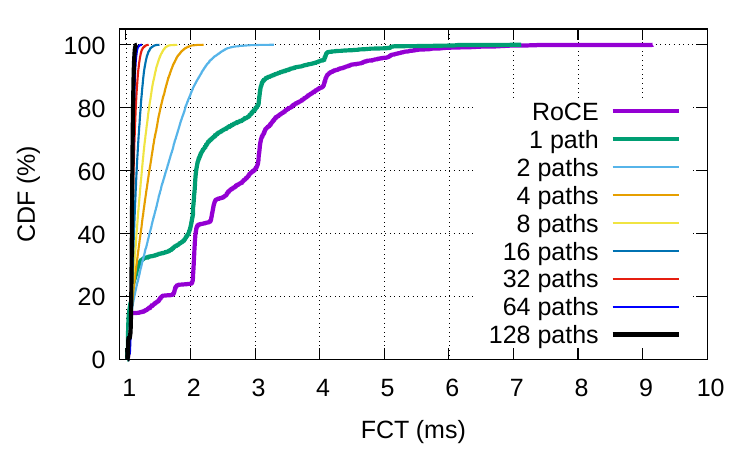}
    \caption{FCT for pipeline parallelism}
    \label{fig:800g_pp}
  \end{minipage}
  \hfill
  \begin{minipage}{0.33\textwidth}
    \includegraphics[width=\linewidth]{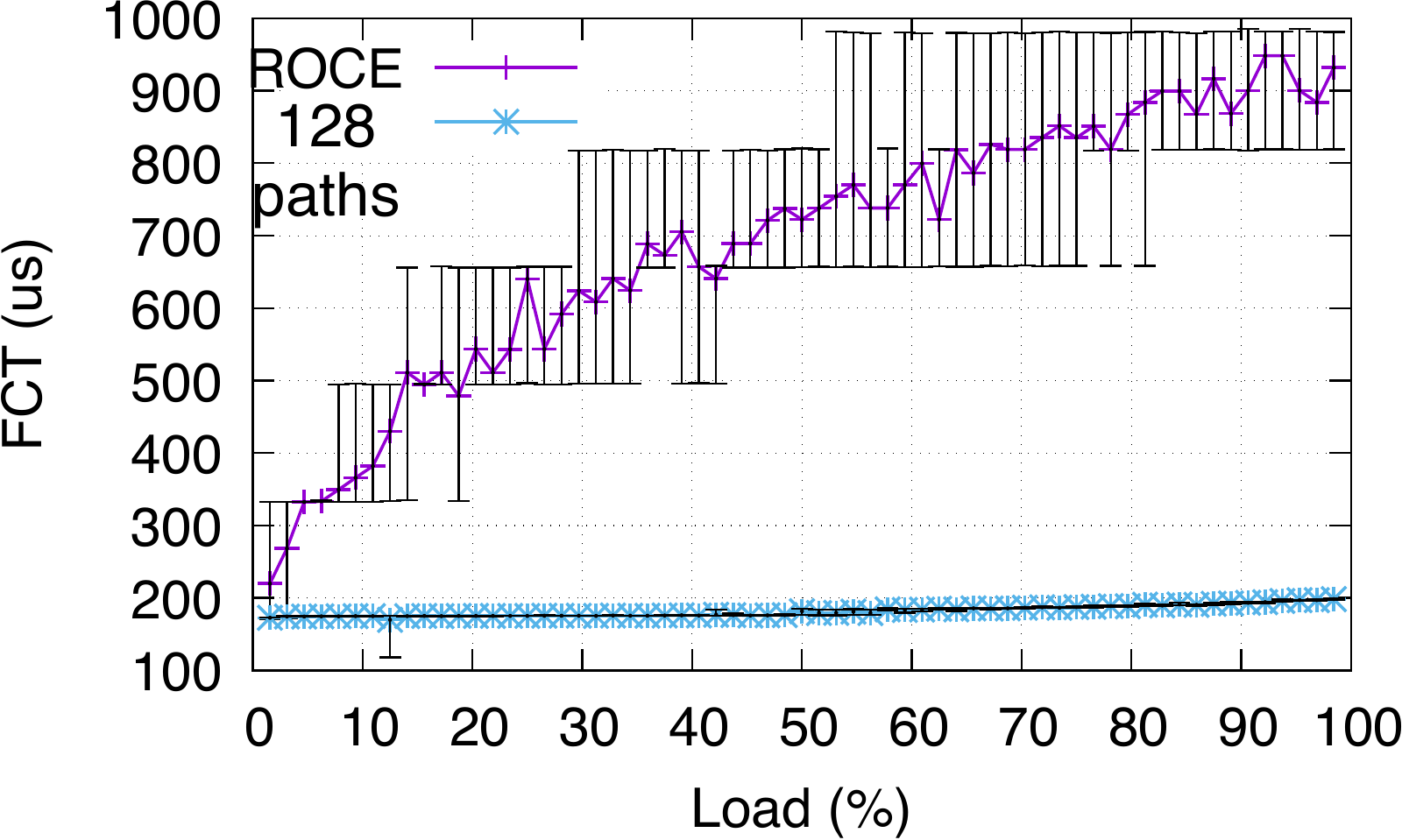}
    \caption{FCT (2MB flows) vs. load}
    \label{fig:perm_load}
  \end{minipage}
\end{figure*}

What if we connect only one GPU from each rack to a separate scale-out network?
The resulting network would have just 11K GPUs, totaling 72
independent networks, or 72 rails (see Figure~\ref{fig:topologies}). Assuming 64
port switches, we would still need three tiers for this network to interconnect
all GPUs. However, if we use a 4x 200~Gbps multi-plane approach, two switch
tiers are sufficient to interconnect the entire datacenter.

\researchquestion{What are the implications of multi-rail, multi-plane
networks to job scheduling (e.g. Slurm) and fault tolerance? (2) How much
further can the costs of the scale-out network be reduced by using traffic
locality? (3) Given a target model, what is the optimal combination of
multi-plane and multi-rail topologies?}

The downside of a full multi-rail approach is that GPU N from one rack can
only talk to GPU N from other racks. Any communication to other GPUs in remote
racks must involve the source or destination rack scale-up network. In the case
of GPU failures, this creates obvious issues. It thus makes sense to reduce the
number of rails to allow more freedom in task placement, as long as it does not
increase the number of tiers.

Taken together, multi-rail and multi-plane can greatly reduce network costs. For
instance, using 76 rails and 4 planes requires 39K switching chips and around
800K switch-to-switch cables per DC. This is a 50\% reduction in the
cost of the switching chips and 66\% reduction in cost of links.
Finally, provisioning the topology to be full bisection does not make sense
since the traffic has locality, and this locality can be exploited by the
cluster manager. This enables further savings.

\subsection{Building the scale-up network}

Current scale-up networks are either proprietary (for instance NVLink / NVSwitch
for Nvidia~\cite{nvswitch}, or InfinityFabric~\cite{infinityfabric} for AMD) or
rely on Ethernet (e.g. Intel's Gaudi~\cite{gaudi}). A recent industry
consortium was formed to standardize scale-up networking~\cite{ualink}.
Interconnect architectures have evolved from mesh interconnects between small
numbers of accelerators (e.g. NVLink, Gaudi2) to switch-based (e.g. NVSwitch for
the Blackwell). In particular, the NVSwitch approach has each GPU connect to
many switching chips, creating a single tier multi-plane
network~\cite{ahead,nvswitch}.

We also note that the workload has evolved from cache-coherent communication
(each GPU has access to the entire memory of the node) to using collectives instead,
making the scale-up network look like a regular network.

Given the lack of standards and growing bandwidth needs, there are many open
research questions in this space.

\researchquestion {What is the right (1) interconnect topology, (2) networking
layer, (3) transport protocol, (4) congestion control and (5) API for scale-up?
(6) Are in-network collective implementations useful (e.g. Nvidia Sharp) or are
end-to-end approaches sufficient as models grow in size?}



\subsection{Transport protocols}
A multi-plane, multi-rail and oversubscribed topology for scale-out is
sufficient on paper, but what transports shall we use? The only
choice today is to use transports based on RDMA, either IP (RoCEv2) or
InfiniBand. Both are lossless and single-path, with similar performance for AIML
traffic (the latter has slightly lower latency). The industry has been moving
towards standardizing an Ethernet transport, to ensure there are multiple sources
for networking hardware~\cite{uec}.


How far are existing transports from the optimal transfer times? Point-to-point
traffic pattern is particularly tough, as it goes across multiple switch tiers. 
We simulated synchronized P2P transfers required for our 100T model using RoCEv2
in a fully-provisioned two-tier FatTree topology with 8192 hosts and 800~Gbps
links. The optimal Flow completion Time (FCT) is around $1~ms$, yet RoCEv2 takes
$9~ms$ to finish, as shown in figure~\ref{fig:800g_pp}. This is due to
collisions because of ECMP~\cite{hedera,mptcp-dc,ndp} and due to head of line
blocking due to PFC.

Lossless networks are notoriously difficult to manage~\cite{microsoft-rdma};
buffers are getting smaller in BDPs as switches are getting faster, meaning
that PFC will trigger more often. Furthermore, congestion control for RDMA such
as DCQCN~\cite{dcqcn} or HPCC~\cite{li2019hpcc} starts at line rate and only
reacts to congestion after the 1st RTT, meaning it can still cause PFC. Indeed,
the industry is moving towards best-effort operation~\cite{uec}.

If we use a state of the art congestion control protocol~\cite{nscc}, best
effort networks and single path, the FCT drops to $7~ms$. This increases
exposed networking from 5\% / 20\% with an ideal transport to 8\% / 25\%. If we
use transports such as NDP~\cite{ndp} or Homa~\cite{homa} that spread traffic
across all paths we get within 5\% of the optimal FCT, as shown in
figure~\ref{fig:800g_pp}. Note that collisions affect FCT even when the network
load is smaller, and do not depend on flow size or the linkspeed;
figure~\ref{fig:perm_load} shows the the tail FCT for 2~MB flow permutation over
a 100~Gbps network when only a fraction of the nodes participate.

\researchquestion{What is the right design for a multipath transport protocol
that can load balance effectively in the presence of failures, can handle packet
losses and is implementable in hardware at 800~Gbps+ speeds? What is the right
split of functionality between switches and the end-hosts with regards to
spraying?}

Finally, the strongest argument for multipath is when multi-plane designs are
used: single path transports require application changes (typically the
collective library) that manually break up the data across the planes, resulting
in sub-optimal load balancing. Furthermore, within a rail, we would have similar
FCT inflation due to collisions as in our experiment above. Indeed, the industry
is adopting multipath transmission and lossy operation as the way
forward~\cite{uec}. 
However, existing RoCE hardware implements go-back-N and
requires PFC to operate well, so big changes are needed.

\headline{Wide area transport.} The scale-out transport must be multipath and
capable of running at 800~Gbps over small RTT links; what about the inter-DC
transport? The all-reduce involves all training nodes, each transferring
\textasciitilde300~MB per iteration in around $100~ms$. This requires at least
5~Gbps per flow. We can use TCP, but the CWND adaptation creates unnecessary
loss; coupled with a fairly large RTT, retransmissions create unnecessarily long
tail FCT which increases exposed networking (figure~\ref{fig:distributed}). If we
adopt a controller similar to B4~\cite{b4}, we can perform both traffic
engineering and admission control on the wide area links to ensure congestion
control is not needed and all traffic is admittable, but this adds further
synchronization issues to be solved. Even with perfect scheduling we still incur
loss due to bit errors, which require retransmissions. These could be avoided by
adding redundancy to the outgoing traffic.
Finally, the wide-area traffic must be prioritized within data center networks,
which means that multipath traffic must be able to cope with non-equal capacity
links. 

\researchquestion{How should a wide-area transport be implemented to enable
near-perfect operation yet be simple to implement and easy to debug? What are
the interactions of wide-area traffic and sprayed traffic?}

\section{Summary}

Significant breakthroughs are required to keep up with the
ever-growing demands of next-generation LLMs. Based on public
information on datacenter building plans from Microsoft and OpenAI, estimations
of model sizes and industry shifts, as well as existing research and
practices, we are able to glean into the challenges that building next
generation AIML training datacenters will bring.

The hybrid network of scale-up and scale-out is replacing the
traditional Clos cloud network, while multiple-planes and
multiple-rails will further complicate communication in ML datacenters. This means
that some nodes cannot directly talk to each other, and that traffic must be split across
multiple planes.
We (and the industry ~\cite{uec}) believe multipath communication is
necessary to make best use of the hardware and mitigate the inflation
single path protocols bring to tail flow completion times. Scheduling
and placement also need to evolve to account for these new networks,
and things are further complicated by failure domains for tensor
parallelism being rigidly confined to the rack level.  Adding wide
area transport into the mix comes with its own set of unique
challenges. It's of utmost importance to leave headroom for redundancy
and failure recovery while keeping utilization high; for example, by
mixing long running training workloads with transient inference
jobs.

It's also important to note that advances in network monitoring
are required to cope with multiple networks (multiple rails and planes for scale-out plus
scale-up) and blazingly fast data speeds, as
current sampling, single network approaches appear to no longer capture the relevant
bigger picture.


\paragraph{\textbf{Acknowledgments}}
We thank the reviewers for their constructive feedback. This work is supported
by VMware gift funding.

\bibliographystyle{ACM-Reference-Format}
\bibliography{ref} 

\end{document}